\documentstyle[aps,prl,multicol,epsf,psfig,boxedminipage,citesort,graphicx]{revtex}

\newcommand{\DIR}{.}

\begin{document}

\twocolumn[\hsize\textwidth\columnwidth\hsize\csname@twocolumnfalse%
\endcsname

\title{Single-vehicle data of highway traffic: microscopic description
of traffic phases}

\author{Wolfgang Knospe$^1$, Ludger Santen$^2$, Andreas Schadschneider$^3$ 
and Michael Schreckenberg$^1$}

\address{$^1$ Theoretische Physik, Fakult\"at 4,
          Gerhard-Mercator-Universit\"at Duisburg,
                  Lotharstr. 1, 47048 Duisburg, Germany}

\address{$^2$ Fachrichtung Theoretische Physik, Universit\"at des
Saarlandes, Postfach 151150, 66041 Saarbr\"ucken, Germany}

\address{$^3$ Institut f\"ur Theoretische Physik,
                 Universit\"at zu K\"oln,
                 Z\"ulpicher Str. 77, 50937 K\"oln, Germany }

\date{\today}

\maketitle

\begin{abstract}

We present a detailed analysis of single-vehicle data which sheds 
some light on the microscopic interaction of the vehicles. Besides the
analysis of free flow and synchronized traffic 
the data sets especially provide information about wide jams which
persist for a long time. The data have been collected at a
location far away from ramps 
and in the absence of speed limits
which
allows a comparison with idealized traffic simulations.
We also resolve some open questions concerning the time-headway 
distribution.

\end{abstract}

\bigskip
]


\section{Introduction}

Like in every other field of physics the modeling of traffic flow is
based on the analysis of empirical data.
In the last few years, several empirical studies were carried out in
order to obtain a 
more profound picture of highway traffic. Many modeling
approaches have been suggested which try to capture the essential
parts of the human driving behavior
(see~\cite{Helbing2000,Chowdhury2000} for an overview). 
On the one hand,
the empirical observation of highway traffic helps to identify and
to characterize traffic phases and the transitions between them. On the
other hand, the theoretical modeling allows simulations of 
large traffic networks and opens perspectives for future
applications like traffic forecasts and dynamic route guidance
systems. However, realistic model approaches of highway traffic have
to be based on empirical results and especially the microscopic
modeling of traffic requires information on the driving behavior of
the vehicles in the various traffic states.

Much progress has been made in the last few years in analyzing empirical
data. Moreover, the technical development of the measurement
techniques and the improved equipment of roads with detectors allows
to observe spatio-temporal structures~\cite{kerner2001} and even
to analyze single-vehicle data~\cite{neubert99,Tilch99TGF}.
The analysis of large sets of minute averaged traffic data has revealed
the existence of 
three traffic phases, namely (i) free flow, (ii) synchronized traffic
and (iii) wide moving jams. This three state picture is nowadays 
widely accepted, although other interpretations of empirical
data have been made~\cite{Helbing99a,Lee2000,Treiber2000}.

Free flow traffic can be characterized by a large average velocity
while congested traffic is simply the opposite of
free flow. 
An important property of congested traffic is that the average
velocity of the different lanes 
can be synchronized~\cite{koshi1983,kerner1996a} leading to large correlations
between the velocity measurements on neighboring lanes. 
Congested traffic can be divided into synchronized
traffic and wide moving jams.

In synchronized traffic, which is mainly observed at on- and
off-ramps, the velocity is considerably smaller than in free flow but
the flow can still have large values. Moreover, the large variance in
density and flow measurements cannot be described by a functional
relationship which becomes visible in an irregular pattern in the
fundamental diagram. Therefore, the cross-correlation between 
flow and density vanishes and thus allows to
identify the synchronized state~\cite{neubert99}. 
In \cite{kernerprl1998} the following picture of the synchronized state 
and a possible jam formation was developed:
While the downstream front of the synchronized traffic region
is usually fixed at an on- or off-ramp the upstream front is separated
from free flow by the so-called pinch region. 
In the pinch region narrow jams are formed. 
In contrast to wide jams, narrow jams consist of only two
fronts where the vehicle speed and the density sharply changes whereas
the width of the fronts of wide jams is 
negligible compared to the bulk region between them. These
jams can merge and form wide jams.

Wide (moving) jams are regions with a very high density and
negligible average velocity and flow.
The width of these structures is much larger than its fronts at the
upstream and downstream ends where the speed of vehicles changes
sharply. A wide jam can be characterized by a few parameters, e.g.,
the velocity with which it moves upstream.
The velocity of the downstream front and the corresponding flow rate
is only determined by the density inside a wide jam and the delay-time
between two vehicles leaving the jam~\cite{Kerner1996}.
In contrast to synchronized traffic wide jams can propagate
undisturbed through either free flow and synchronized traffic without
impact on these states which thus allows their coexistence. 

In order to identify the different phases it would be ideal to
have data from a series of detectors in order to identify the
traffic states by their spatio-temporal properties~\cite{kerner2001}.
Since the data set used in this study is recorded at a single
measurement location we use autocorrelation and cross-correlation
functions to determine these properties locally.

The analysis of single-vehicle data gives
important information about the driving behavior of the vehicles in the
various traffic states. 
Unfortunately, up to now only a few empirical results of
single-vehicle data exist~\cite{neubert99,Tilch99TGF} which do not provide a
consistent picture of the vehicle dynamics. 
It was shown that the distribution of the time-headways
and the velocity-distance relationship do not depend on the density
but on the traffic state. In~\cite{neubert99} a peak at $1.8$ s was
found in the time-headway distribution which was explained with the
drivers effort for safety. However, the existence of this peak could
not be confirmed in~\cite{Tilch99TGF}.
Moreover, in the free flow state platoons of vehicles driving
bumper-to-bumper with a time-headway well below $1$ s occur.

The velocity-distance relationship gives information on the adjustment
of the velocity on the headway. 
While in free flow the velocity shows large values even for small
headways, in the congested state drivers tend
to move slower than the distance allows.
Thus, in the free flow regime the movement of the {vehicles} is
uncorrelated although small platoons exist while in the synchronized
regime large correlations of the velocity between the vehicles can be found
but the distance between them can vary considerably.

In this empirical analysis we try to improve the microscopic understanding
of the human driving behavior which helps to validate the properties
of models for traffic flow on a qualitative level. 
It is worth pointing out that the data set underlying our study is one
of the largest used so far. In contrast to older studies the data
have been collected on a 3-lane road without speed limit and
relatively (several kilometers)
far away from ramps. The quality of the
empirical data enables a detailed insight into the dynamics of the
vehicles in wide jams. 
The location of the detectors far away from ramps and the absence of a
speed limit allow a much better comparison with data from idealized
traffic simulations without bottlenecks.
In analogy to onramps~\cite{Diedrich2000}, defects introduced by,
e.g., the application of a speed limit, considerably affect traffic 
flow and can trigger distinct traffic states. However, it has been 
shown recently for a large class of driven diffusive systems 
(see e.g.~\cite{Knospe01,popkov99}) that bottlenecks only select rather 
than generate the states. 
This is a very general property of driven system and is well-established
in nonequilibrium statistical physics (see ref.~\cite{schuetz} and
references therein). 
Furthermore it has also been verified for a large class of traffic
models.
Therefore our analysis has, compared to 
measurements at bottlenecks, the advantage that we can observe 
and analyze pure bulk states. The disadvantage is that congested traffic 
is less often observed, which implies that there is need for 
quite long observation periods. However, the data set, which is 
analyzed throughout this article, includes all three types of 
highway traffic.
Our results are therefore of direct relevance for the
modeling of traffic flow on highways.


\section{Empirical data}
\label{occupancy}

\begin{center}
\begin{figure} 
\includegraphics[width=\linewidth,height=3cm]{\DIR/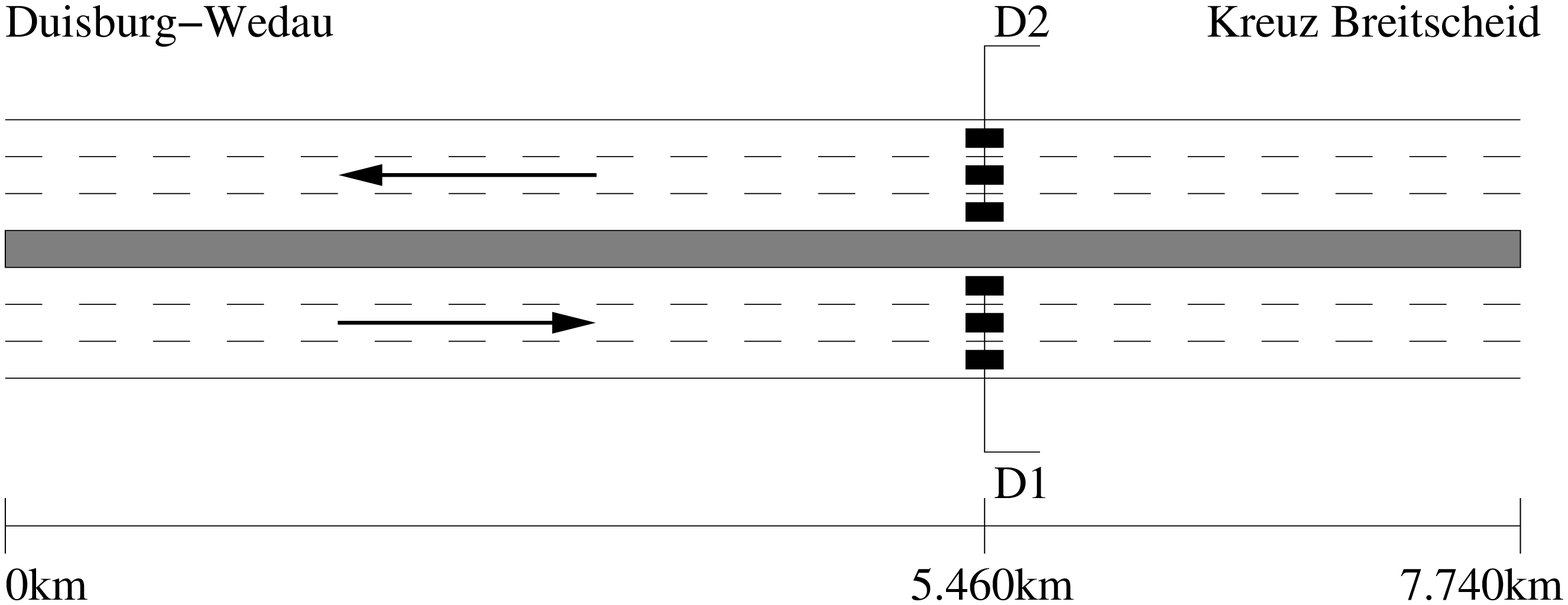}
\includegraphics[width=\linewidth,height=3cm]{\DIR/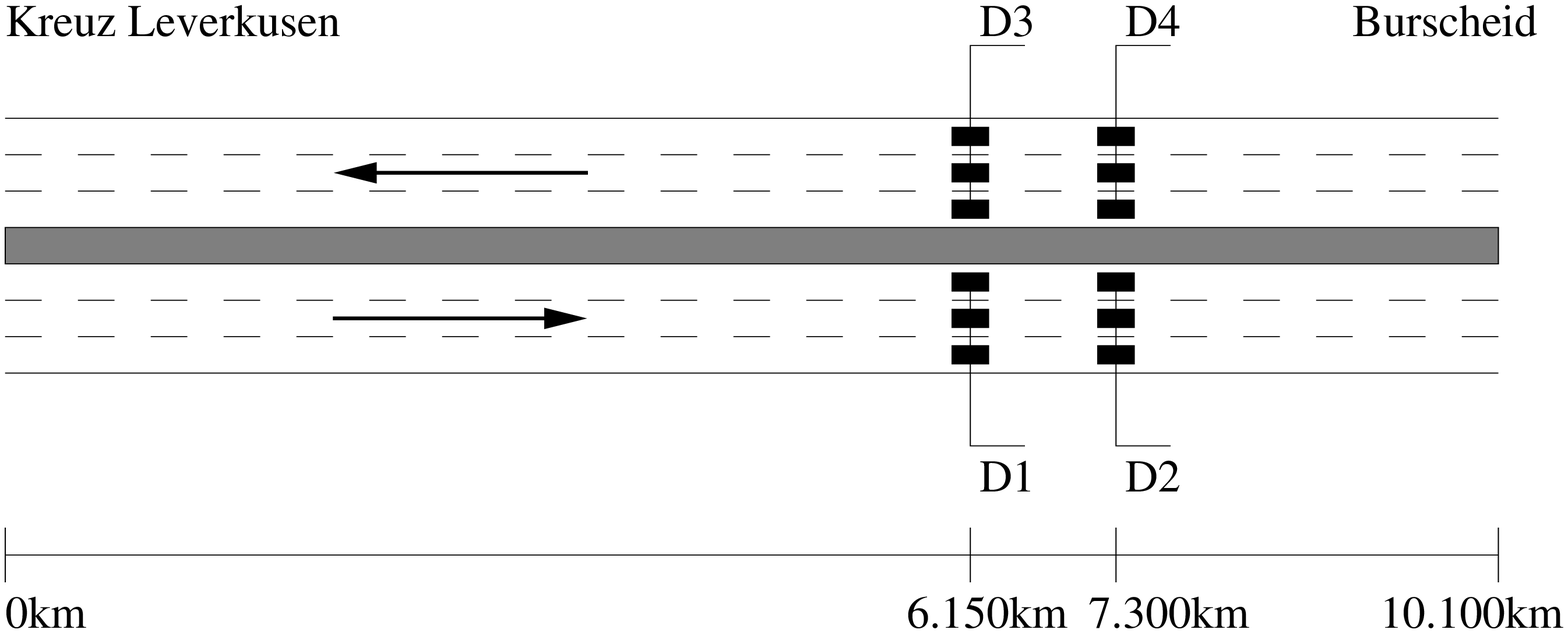}
\caption{Location of the detectors of the first data (top) set measured on
the A3 near Duisburg and of the second
and third data set (bottom) measured on the A1 near Leverkusen.}
\label{fig_a3}
\label{fig_a1}
\end{figure}
\end{center}

The data sets were collected at different locations on two highways in
North-Rhine-Westfalia, Germany. The first set is provided by two
detectors, one for each driving direction, on the A3 between the
junction 
Duisburg-Wedau and the highway-intersection Kreuz-Breitscheid (see
Fig.~\ref{fig_a3}). A detector consists of three inductive loops, one
for each lane. The location of the detector is about $5.5$ km from the
highway intersection Breitscheid and about $2.2$ km from the junction
Duisburg-Wedau. It is important to note that there is no speed limit.
The data set was measured between 03-30-2000 and 05-16-2000 and
comprises $48$ days.
The second and the third data set was measured by four detectors, 
one for each driving direction 
consisting of three inductive loops, at two locations on the A1
between the highway intersection Leverkusen and the junction
Burscheid (see Fig.~\ref{fig_a1}). 
Note, that for storage reasons, in the third data set only the vehicles on the
right and the middle lane were detected.
The detectors are located
$6.150$ km, respectively $7.300$ km from Leverkusen and $3.950$ km,
respectively $2.800$ km from Burscheid with a distance of $1.150$ km
between the 
two measurement locations. 
The data were collected during two periods (08-02-2000 to 08-20-2000 and
02-01-2001 to 02-27-2001) which comprises a total measurement time of
$46$ days.
Note, that the highway has three lanes and there is no speed limit.
In total, on $94$ days about $9.5$ million cars were measured.

All of the inductive loops were able to measure single-vehicle data.
A data set of one vehicle comprises the time (in hundredth seconds) at
which the
vehicle reaches the detector, the lane the car is detected, the type of
the vehicle (car, truck), the velocity in km/h and its
length in cm. The length of a vehicle is calculated with
an accuracy 
of $1$ cm. The lower bound for the velocity measurement is
$10$ km/h, i.e., velocities of slower vehicles are not measured.
Then, instead of the length of the vehicle, the time the inductive loop 
is covered is given in hundredth seconds. 
Note, that the velocity is theoretically calculated internally via the
time $\Delta t$ a vehicle needs to pass a detector of a given length
with an accuracy of $1$ km/h. 
However, since $\Delta t$ is only measured in milliseconds and no
floating points for the calculation of the velocity are used, 
it is possible that a velocity may correspond to more than one $\Delta
t$ and/or a difference of one millisecond in $\Delta t$ leads to
a velocity difference larger than $1$ km/h.
As a result, small differences in $\Delta t$ cannot be resolved while
some velocities cannot be measured.

The third data set includes the electric signal inducted in the loops 
when a car passes the detector. These data comprise the time (in
milliseconds) the vehicle arrives at the detector and the time-series
of the height of the
electric signal that is sampled at 125 points in one second.

In addition to the single-vehicle data of the first and the second
data set, one minute averages of the
traffic data are collected. These data comprise the number of cars and
the number of trucks, the mean velocity of cars, the mean velocity
of trucks and the mean velocity of all vehicles. 
Occupancy ($occ$) rather than density is used as an estimation of the
density because of the ease of measuring and its independence from the
vehicle lengths.
It is simply given by the fraction of time the detector is covered
by a vehicle.

The spatial density can be calculated from the occupancy.
An occupancy $occ=1$ means that the vehicles are driving
bumper-to-bumper which leads to a spatial density $\rho_{max}$ of 
\begin{equation}
\rho_{max}
= \frac{N(T)}{\sum_{t \le t_n < t+T} l_n} = \frac{1}{L}
\end{equation} with the mean
length $L$ of the vehicles and the number $N(T)$ of the vehicles
passing the detector in time $T$. In this study the value of $T$ is set to
$60$ s. $l_n$ is the length of the $n$-th
vehicle measured at time $t_n$.
For the spatial density $\rho_s$ estimated by occupancy 
\begin{equation}
\rho_s = occ \cdot \rho_{max} = occ \cdot
\frac{N(T)}{\sum_{t \le t_n < t+T} l_n} = \frac{occ}{L} 
\label{occden}
\end{equation}
holds~\footnote{The calculation of the spatial density by means of
local measurements is only valid under the assumption of a constant
velocity of the vehicles during the averaging process.}.
Throughout this paper the terminology ``density'' is used for 
occupancy if not stated otherwise and the occupancy is given in percent.

Since large occupancies (e.g., a vehicle blocks the detector for more
than one minute) are not distributed on minute
intervals by the counting device but the occupancy is just related to
the interval the 
vehicle is measured, a cut-off of the minute averaged data at $100~\%$
was introduced. 
Thus, the explicit calculation of the occupancy of the time interval
[t,t+T] by means of the
single-vehicle data via  
\begin{equation}
 occ(t) = \frac{1}{T} \sum_{t \le t_n < t+T} \Delta t_n =
\frac{1}{T} \sum_{t \le t_n < t+T} \frac{l_n}{v_n} 
\end{equation}
with the time $\Delta t_n$ the detector is occupied by car $n$ of length
$l_n$ and velocity $v_n$
allows to determine
the occupancy more accurately compared to the one minute data because
occupancy overlaps from minute to minute can be considered.
The minute data therefore are only used to verify the averaged 
single-vehicle data.

Furthermore, occupancy can explicitly be related to the various traffic
states. 
In contrast, the hydrodynamical relation 
\begin{equation}
\rho = \frac{J}{v}
\end{equation}
where $J$ is the flow (which is proportional to the number $N$ of
vehicles passing the detector in a time interval $\Delta t$) and $v =
\sum v_n(t)/N$ is the average velocity of the vehicles tends to
underestimate the density since the velocity of {\it moving} cars is
detected (see~\cite{neubert99}). 
This leads especially in the congested state to
very small densities while in free flow traffic both methods give the
same results.

The time resolution of the single-vehicle data allows the
calculation of the time-headway $t_{h}$ and the distance-headway $gap$
of the $n$-th vehicle via 
\begin{equation}
t_{h}(n) = t_n-t_{n-1}-\frac{l_{n-1}}{v_{n-1}}
\end{equation} and 
\begin{equation}
gap(n) = v_n
(t_n-t_{n-1}) - l_{n-1}
\end{equation}
under the assumption\footnote{The usage of the velocity $v_n$ is 
somehow arbitrary since $v_{n-1}$ can also be taken which leads to 
the calculation of the headway in upstream direction rather than downstream.
However, the difference between both results is negligible because 
on small distances the surroundings upstream and downstream of the
measurement location should be homogeneous.}
of a constant $v_n$ and $v_{n-1}$.  
$t_n$ denotes the time the $n$-th vehicle 
passes the detector, $l_n$ and $v_n$ its length and velocity.
The passing time  are measured by the internal clock with 
an accuracy of  $0.001$ s. In the output these signals are 
provided in units of  hundredth seconds. The achievable resolution 
of spatial measurements is $1$ cm. These values are of course 
lower bounds for the real error bars. In particular for the distance 
headways we expect much uncertainties, roughly of the order of $\sim
1$ m. 
Unfortunately, it is difficult to calculate precisely one realistic errorbar,
because they depend on, e.g., the velocity of the cars and the actual 
value of the distance.

\begin{center}
\begin{figure} 
\includegraphics[width=0.9\linewidth]{\DIR/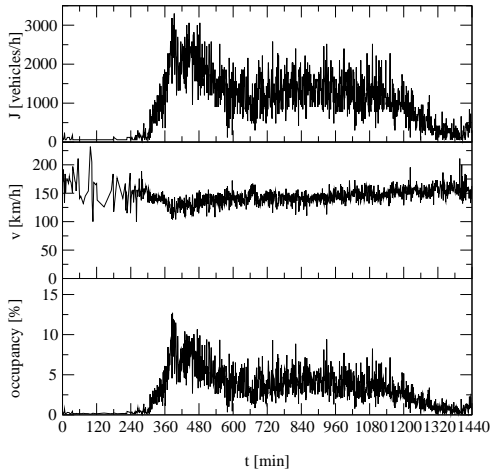}
\caption{Time-series in the free flow regime of the flow, the
velocity and the occupancy on the left lane at detector D1 on the A3 on the
04-28-2000.}
\label{ff_fundi1}
\end{figure}
\end{center}

In order to allow the calculation of spatio-temporal correlations on the A1,
the two measurement locations D3/D4 and D2/D4 are synchronized by
means of radio controlled clocks.


\section{Free flow}

Free flow is the traffic state most frequently observed in the data set and
is characterized by a large flow $J$ and velocity $v$ and a small
density (Fig.~\ref{ff_fundi1}). 
While daily variations of the flow and the density become visible in
the time-series\footnote{Quantitatively they can be seen in the
behavior of the autocorrelation functions \cite{neubert99}.}, the
velocity remains nearly constant over the whole day.

\begin{center}
\begin{figure}
\includegraphics[width=0.9\linewidth]{\DIR/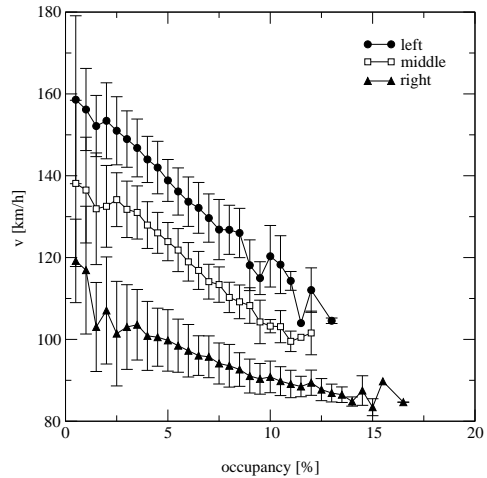}
\caption{Average velocity dependent on the density for the three lanes
at detector D1 on the A3 on the 04-28-2000.}
\label{v_mean}
\end{figure}
\end{center}

Due to legal restrictions in
Germany vehicles have to drive on the right lane and the left and
the middle lane should be used only for 
overtake maneuvers. Thus, the lanes segregate into fast and slow lanes which
results in a negative velocity gradient from left 
to right (Fig.~\ref{v_mean}). This effect becomes more pronounced in
the absence of a 
speed limit. 
As a consequence, 
the flow and the velocity decreases from the left to the right lane,
whereas the density increases.

\begin{center}
\begin{figure}
\includegraphics[width=0.9\linewidth]{\DIR/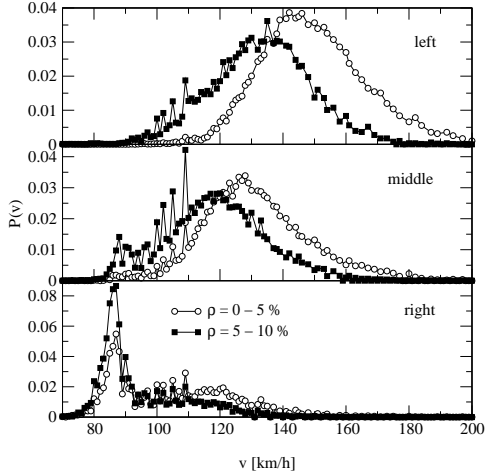}
\caption{Distribution of the velocity on the three lanes at
detector D1 on the A3 on the 04-28-2000.}
\label{ff_vdis}
\end{figure}
\end{center}

Next we analyze the velocity distributions in free flow traffic
(Fig.~\ref{ff_vdis}). 
The velocity calculated by the detector is simply the length of 
the detector divided by the time $\Delta t$ the vehicle takes to pass the
detector, that is measured directly.  
In our case $\Delta t$ is measured with a precision of milliseconds, but the 
corresponding velocity is given in  
units of km/h since only integer arithmetics was applied
internally. This way of calculating the velocity does not  
use the full precision of the measurement, but even worse, it leads to an 
uneven sampling of the velocities. The latter point can be avoided 
by recovering the corresponding traveling time $\Delta t$ from the velocity 
given in the data set. For large velocities ($v> 110$ km/h) at one hand
we obtain a unique value of $\Delta t$.
For smaller velocities on the other hand each velocity corresponds 
to a range of possible 
traveling times. If the same velocity results for $k$ different 
traveling times, we simply increase the statistical weight of each possible 
value of $\Delta t$ by $1/k$. This procedure leads to smooth 
traveling time distributions. {\bf If one is interested in the 
velocity distributions one is still left with a problem, i.e. 
the velocity intervals are that correspond to a given traveling time 
are not of equal length. Therefore we calculated the velocity, 
that corresponds to a certain travelling time, with a high
numerical resolution, i.e. $10^{-2}$ km/h\footnote{The final distribution 
function does not change if a higher resolution is chosen.},
 and interpolated linearly between the two measurements. In order 
to obtain a normalized distribution the interpolated values have to be 
devided by the length of the interval. In the final figure, however,
we use a resolution of $1$ km/h, by averaging over the evenly sampled 
and highly resolved velocity distribution. This procedure leads
 obviously to very smooth velocity distributions.}

\begin{center}
\begin{figure}
\includegraphics[width=0.9\linewidth]{\DIR/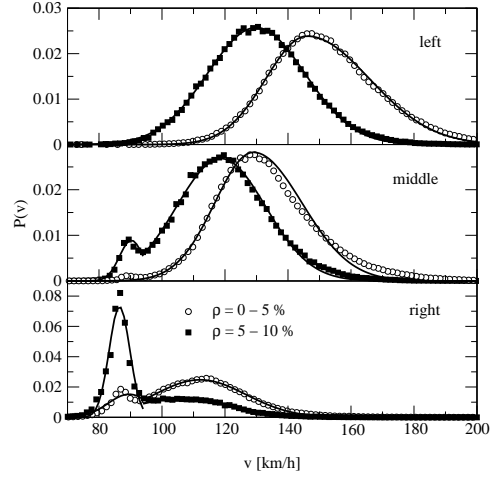}
\caption{Distribution of the recalculated velocity (see text).
The solid lines represents best fits with distributions of Gaussian
type. For distribution functions that have two maxima a superposition
of two gaussian distributions has been used. For single peaked 
distributions a simple or, in case of asymmetric distributions, 
a distribution consisting of two piecewise defined Gaussians has been
used as a fit  function. (see table~\ref{tab}).} 
\label{ff_vdis_ok}
\end{figure}
\end{center}

On the left lane the velocity distributions can be nicely fitted 
to a Gaussian distribution, for densities between $5-10\%$. This 
is in agreement with earlier investigations, also finding 
Gaussian distributed velocities, for a low portion of
trucks~\cite{helbing1997}.  
For lower densities ($0-5\%$), however, we observe a slight asymmetry of the 
distribution function. In order to quantify this asymmetry we 
divided  the distribution function into two parts, one containing 
all velocities below the maximum of the distribution, and the other 
for larger velocities. Both parts are in agreement with different 
Gaussian distributions (Fig.~\ref{ff_vdis_ok}). 

The form of the distribution function on the middle and right lane 
is different, because they can be used by trucks. The trucks lead to a second
maximum of the velocity distribution. For these two lanes the velocity
distribution can be viewed as a superposition of a ``car''and a ``truck''
velocity distribution, both of Gaussian form in the density interval 
$5-10\%$. For lower densities we 
find again an asymmetry of the distribution function
that is even more pronounced than for the left lane.
For an overview, the fit-parameters are summarized in table~\ref{tab}.

\begin{center}
\begin{figure} 
\includegraphics[width=0.9\linewidth]{\DIR/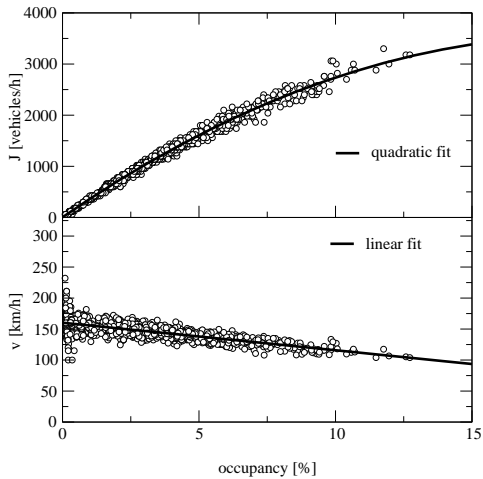}
\caption{Fundamental diagram of the left lane for the flow and the
velocity in 
the free flow regime on the 04-28-2000 at detector D1 on the A3. The
solid curve gives the best fit for the 
data.}
\label{ff_fundi_fit}
\end{figure}
\end{center}

In the presence of a speed limit it was found that the negligible 
vehicle interactions are one of the main characteristics of free flow 
traffic \cite{neubert99}.
In contrast, the
velocity-density relationship in 
Fig.~\ref{ff_fundi_fit} clearly deviates
from a constant value.
This fact can be explained by the absence of a speed limit.

\begin{table}
\begin{tabular}{p{1.1cm}p{3.1cm}p{0.6cm}p{0.7cm}p{0.9cm}}
density & distribution & a & b & c \\
\hline\\
$0-5\%$ &left, $v < 146$ km/h & $146.0$ & $340.79$ & $41.93$\\
&left, $v \ge 146$ km/h & $146.0$ & $789.86$ & $41.93$\\
 &middle, $v < 129$ km/h & $129.0$ & $300.21$ & $35.46$\\
&middle, $v \ge 129$ km/h & $129.0$ & $477.137$ & $35.46$\\
& right, cars & $112.10$ & $424.22$ & $40.40$\\
& right, trucks & $89.55$ & $88.44$ & $65.70$\\\hline

$5-10\%$ &left & $129.9$ & $477.84$ & $38.92$\\
& middle, cars & $118.75$ & $401.50$ & $37.27$\\
& middle, trucks & $90.27$ & $28.10$ & $112.27$\\
& right, cars & $106.24$ & $466.06$ & $80.95$\\
& right, trucks & $86.51$ & $21.72$ & $13.69$\\

\end{tabular}
\caption{Fit parameters of the velocity distribution in free flow. As
fit function the Gaussian $p(v) = \exp(-(v-a)^2/b)/c$ was used.}
\label{tab}
\end{table}

Slow vehicles are responsible for the decrement of the mean velocity
with increasing density. The average velocity
decreases linearly with the occupancy ($v \approx 160.206 \pm 0.4884 -
(4.44036 \pm  0.1169) occ$) and consequently the flow 
quadratically, $J \approx 5.02139 \pm 4.334 + (368.543 \pm 2.177)
occ - (9.54743 \pm 0.2377) occ^2$.

In contrast, queuing theory predicts for a M/G/2 queuing system
that has to be 
applied for a three-lane highway~\cite{May90} a quadratic relationship
between 
density and velocity and a cubic relationship between density and
flow, respectively.  For two-lane highways (a M/G/1 queuing system)
the velocity is expected to depend linearly on the density.

The discrepancy between queuing theory and empirical data may 
indicate that either the theory is not appropriate or that the three 
lanes are not homogeneously coupled. The latter point of view is 
supported by a correlation analysis of the different lanes, showing 
that the coupling between left and middle lane is much stronger 
than between the right and middle lane. It would be interesting 
to parameterize the partial decoupling of the right lane and 
to study the modified queuing theory. 

\begin{center}
\begin{figure}[hbt]
\includegraphics[width=0.9\linewidth]{\DIR/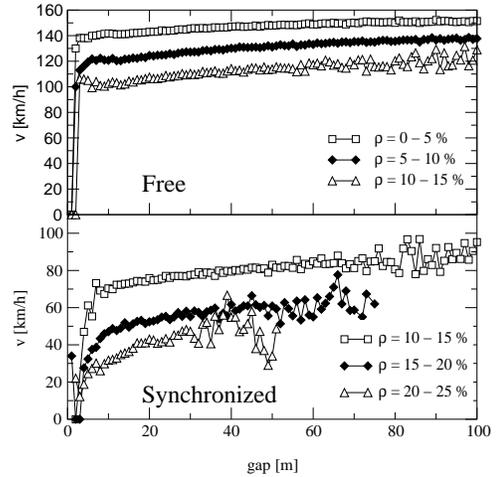}
\caption{The mean velocity
chosen for a certain distance-headway 
at detector D1 on the left lane of the A3 for all days with free flow
and synchronized traffic.}
\label{s_ov}
\end{figure}
\end{center}

The velocity-headway curve illustrates the vehicle's velocity
adjustment on the distance-headway. Since cars can move freely, the
velocity saturates even for small distances. However, with increasing
vehicle interaction the asymptotic value of the velocity decreases
(Fig.~\ref{s_ov}).

\begin{center}
\begin{figure} 
\includegraphics[width=0.9\linewidth]{\DIR/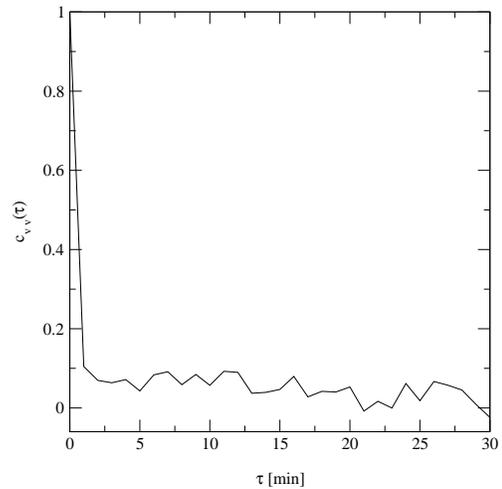}
\caption{Autocorrelation of the minute averaged velocity of the left
lane on the 04-28-2000 at detector D1 on the A3.}
\label{ff_ac1}
\end{figure}
\end{center}

The vehicle interactions, although visible in the
fundamental diagram, are very small.
In Fig.~\ref{ff_ac1} the autocorrelation (see
appendix~\ref{fourier}) 
of the one-minute averages of the velocity 
is shown. 
In order to eliminate the daily variations the time-series of the
velocity was detrended\footnote{The data were fitted using a linear
regression. This trend was then subtracted from the data points.} by a
linear fit.   
Obviously, the minute averaged velocity shows no
correlations on time scales larger than one minute since the vehicles
are in general driving independently.

\begin{center}
\begin{figure} 
\includegraphics[width=0.9\linewidth]{\DIR/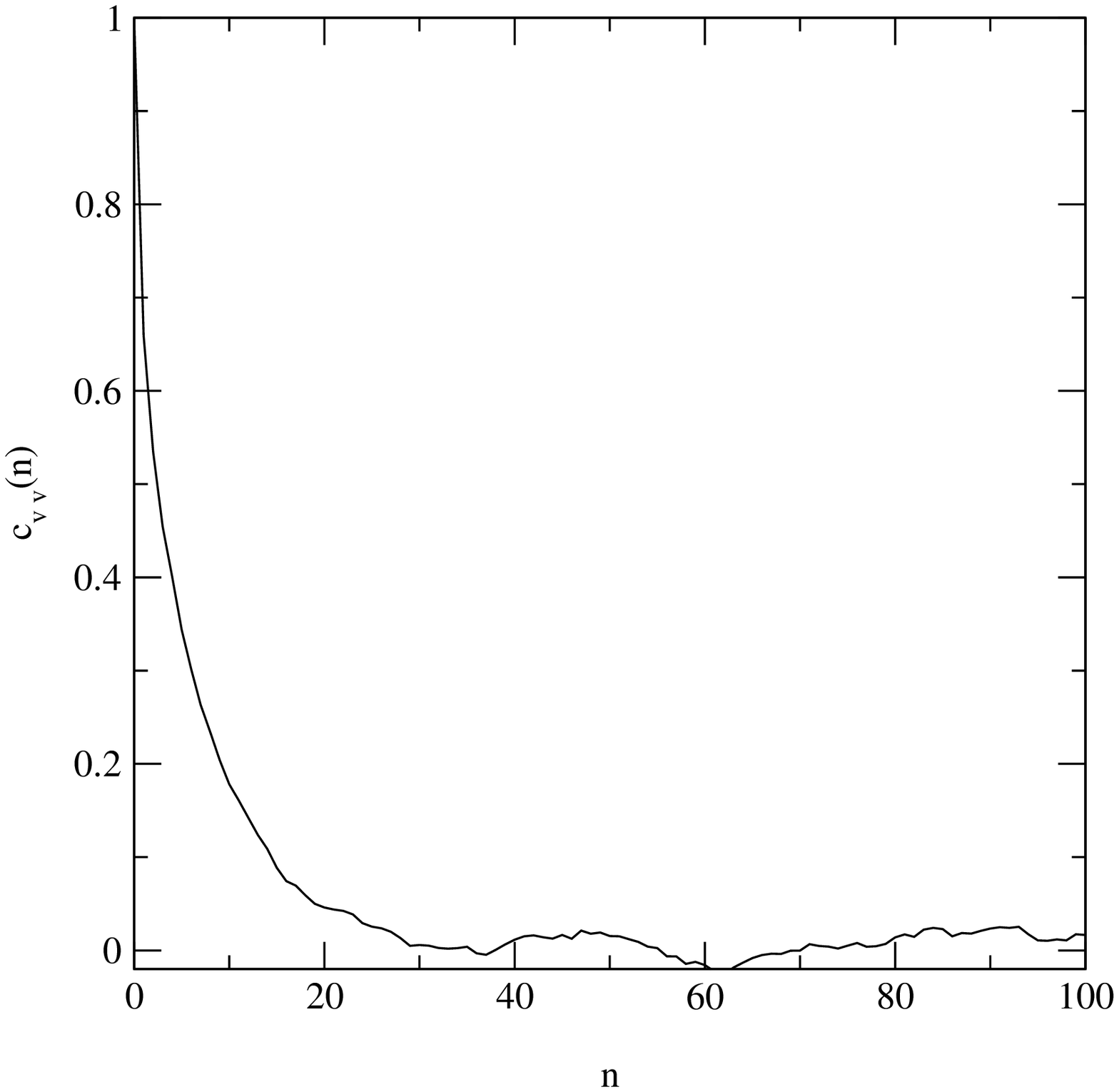}
\caption{Autocorrelation of the single-vehicle velocity in free flow of the
left lane at detector D1 on the 04-28-2000 on the A3.}
\label{ff_ac2}
\end{figure}
\end{center}

\begin{center}
\begin{figure} 
\includegraphics[width=0.9\linewidth]{\DIR/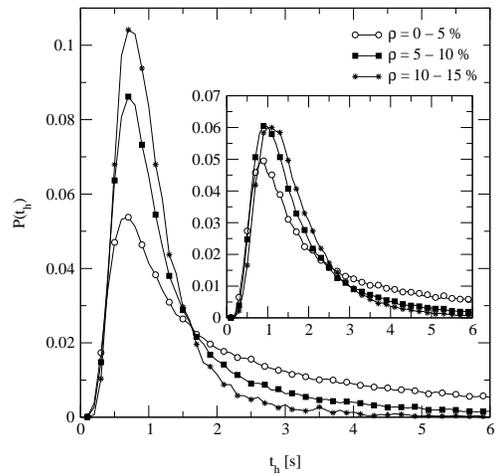}
\caption{Distribution
of the time-headways of the left lane at detector D1 on the A3 for all
days of the data set with free flow. The inset shows the time-headway
distribution of new measurements on the A1 near
K\"oln-L\"ovenich.} 
\label{ff_th1}
\end{figure}
\end{center}

Nevertheless, a closer look on the autocorrelation
of the velocity of the single-vehicle data 
(Fig.~\ref{ff_ac2}) reveals correlations on very short
time-scales.
Like the minute-averaged data, the 
single-vehicle data were detrended to eliminate daily variations.
Now, the autocorrelation function is not time-dependent
but depends on the number of vehicles.
The strong correlations on short time-scales can be attributed to small
platoons of cars driving with nearly the same speed, but with very small
headways~\cite{neubert99}.
The platoons are also responsible for small time-headways in the
time-headway distribution (Fig.~\ref{ff_th1}).
As one can see from Fig.~\ref{ff_th1} the time-headway distribution
differs from the distribution found in~\cite{neubert99}.
In~\cite{neubert99} the time-headway distribution of the
single-vehicle data shows the existence of a peak at $1.8$ s for all
densities and all traffic states which was explained to be the optimal
safe time-headway between vehicles.

This peak was missing in 
related empirical investigations \cite{Tilch99TGF} including the present.  
In order to clarify the origin of this peak we analyzed the data used 
in~\cite{neubert99} again. A detailed analysis of this data set revealed
even more than one peak, but also smaller peaks for other multiples of 
$0.9$ s. Therefore one is let to the believe that these peaks could be
artificially  
generated by a software error of the detection device. Indeed it
turned out that  
in case of a busy detector the data are not directly transfered but 
stored in a buffer for a time of $0.9$ s.

Since the single-vehicle data contain no time-stamp and the storage
time of the internal clock of the computer was used, 
to each data point in the buffer the same time of measurement is
assigned.
In the time-headway distribution 
then the multiples of the buffer size become visible. 
This is confirmed by new measurements (see inset of Fig.~\ref{ff_th1})
at the detectors D2 and D3  
on the A1 near K\"oln-L\"ovenich that have also been used
in~\cite{neubert99}. This data set (like the data sets used in this
paper) records the time  a vehicle arrives at the
detector in hundredth seconds which therefore allows the calculation
of the exact time-headway.  
The new data do not show a peak at $1.8$ s in the time-headway 
distribution (about $1.5$ million cars were detected in $19$ days).

At this point we stress the fact that the velocity and
length measurements of the single-vehicle data of~\cite{neubert99} are
not affected by the software error and more important, the order of
the data is not mixed up. In particular, results of the fundamental
diagram, the time-series analysis and the optimal-velocity curve can
be reproduced by the data sets used here.

The time-headway distribution found here, shows a strong dependence on the
density. However, although the width of the distribution decreases the 
short time
behavior remains nearly unchanged. Not only the maximum of the
distribution is at about $0.68$ s, but also the values of the shortest
time-headways do not change significantly. Note, that the distribution
was calculated by considering all days of the data set with
free flow traffic. 

\begin{center}
\begin{figure} 
\includegraphics[width=0.9\linewidth]{\DIR/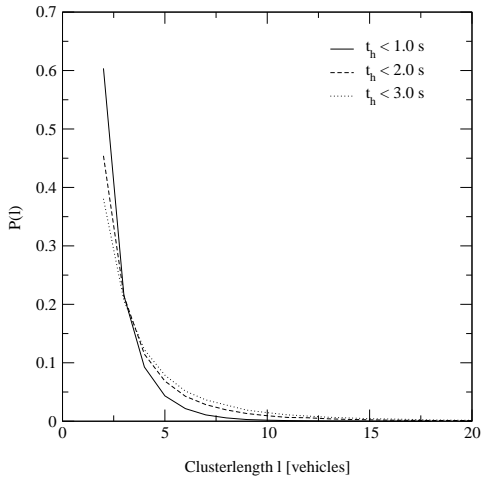}
\caption{Cluster length distribution on the left lane in the free flow
regime for different time-headways.}
\label{ff_cluster1}
\end{figure}
\end{center}

If one compares the time-headway distributions 
of the highway with and without speed limit, one observes that the maximum 
of the distribution is shifted towards larger time-headways in case 
of an applied speed limit. This observation shows the importance of 
anticipation effects in free flow traffic. However, the anticipation
of the predecessor's movement 
leads to serious problems concerning safe 
driving, because the drivers adjust the distance to the leading 
vehicle discarding their own velocity. 

Another important point is the cut-off of the distribution at 
small time-headways. In both cases we found that a non-negligible 
portion of vehicles drives with a temporal distance of the 
order $\sim 0.2$ s, that is comparable to the reaction time of the
drivers. These drivers take a very high risk to provoke an accident
and would definitively be unable to avoid a crash in case of an 
sudden obstacle.

It is now possible to estimate the temporal extension of the
platoons. From the decay of the autocorrelation of the single-vehicle
velocity a 
platoon length of the order of magnitude of about $10$ 
vehicles follows. Since the time-headways within a platoon are smaller than
the maximum of the distribution (about $1$ s) this leads to values
less than one 
minute. Thus, these small vehicle clusters cannot be visible in the
minute averaged data.

The analysis of the single-vehicle data allows to measure cluster
lengths directly.
A platoon of length $l$ consists of $l$ consecutive vehicles each having
a time-headway smaller than a defined value\footnote{This value should be
of the order of the maximum of the time-headway distribution, i.e.,
about $1$ s.}. 
By counting the number of succeeding vehicles with time-headways
well below a certain value one obtains the cluster length
distribution (see Fig.~\ref{ff_cluster1} for the distribution of the
left lane).

\begin{center}
\begin{figure} 
\includegraphics[width=0.9\linewidth]{\DIR/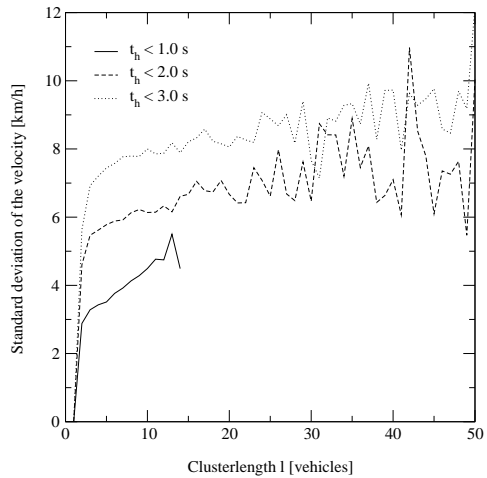}
\caption{Average standard deviation of the mean velocity of the
vehicles in a cluster of length $l$ on the left lane.}
\label{ff_cluster2}
\end{figure}
\end{center}

The length of the platoons increases with increasing time-headway, but
even with very small time-headways lengths up to $14$ vehicles can be
measured. 
In contrast to vehicles on the middle and the right lane, cars on 
the left lane have no possibility to overtake a slower one. 
As the length of the platoons decreases from the left to the right
lane one can conclude that platoons are simply formed by fast 
cars piling up behind a slow car (see~\cite{islam} for a functional
relationship of the bunch size distribution).

Driving bumper-to-bumper with large velocity requires an increased
attention to velocity fluctuations of the preceding vehicle. As a
consequence, the velocity of vehicles inside a platoon 
is almost synchronized which leads to a slow decay of the autocorrelation
function. 
Driving in a synchronized manner also decreases the velocity
differences within the platoon. In Fig.~\ref{ff_cluster2} the standard
deviation of the velocity of all platoon vehicles is depicted. As one
can see, the smaller the time-headway the smaller the velocity
differences. 
Furthermore, the standard deviation saturates very fast so that even
large clusters move as synchronized as small ones.

\begin{center}
\begin{figure} 
\includegraphics[width=0.9\linewidth]{\DIR/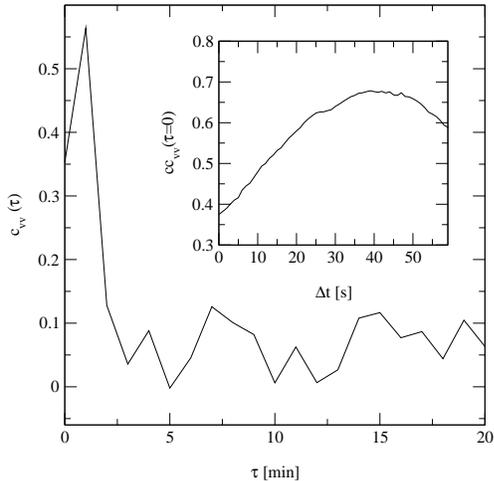}
\caption{Cross-correlation between the minute averaged velocities at the
detectors D3 and D4 of the left lane on the A1 on the 08-10-2000. The
inset shows the cross-correlation for $\tau = 0$ of the velocity
time-series at D4 
and the by $\Delta$ seconds shifted velocity time-series at D3.}
\label{ff_spatialcorr}
\end{figure}
\end{center}

In order to evaluate the stability of the platoons, i.e., identify stable
moving structures, we calculate the cross-correlation between the
minute averaged 
velocities at the detectors D3 and D4 on the A1
(Fig.~\ref{ff_spatialcorr}). If the platoons were 
stable the order of the vehicles passing a detector remains unchanged
leading to a strong correlation of the velocity time-series.
Vehicles on the left lane need about $30$ s to drive from $D3$
to $D4$. As one can see, there is indeed a strong correlation of both
sites at $\tau=1$ min, indicating that the lifetime of the platoons may
have large values and moving structures can be identified even in free
flow. 

The analysis of single-vehicle data allows to specify the lag between
the two time-series by varying the displacement of the two time-series
continuously. The starting point of the
averaging interval of the time-series at D3 is shifted by
$\Delta$ seconds and then, the cross-correlation of
the two signals is calculated. 
If the two time-series match, the cross-correlation would show large
values at $\tau = 0$.
Hence, it is possible to locate the maximal
correlation which corresponds roughly to the average travel time
$\Delta_{\rm max}$ given by 
$1150$ m$/150$ km/h $\approx 28$ s (see inset of
Fig.~\ref{ff_spatialcorr}).

As mentioned before, the third data set provides information about the
signal the 
inductive loop measures from each car. We therefore tried
to identify the moving structures of free flow traffic not only on the
minute averaged scale but also on the single-vehicle
scale. Unfortunately, the
consideration and comparison of the vehicles electric signal at both
detectors fails for two reasons. 
First, the electric signal differs only slightly from car to
car. Second, even on the right lane the sequence of cars changes due
to lane changes and overtaking maneuvers. 
As a consequence, the number of vehicles on one lane is not
conserved so that even the cumulative number of vehicles passing a
measurement section cannot be compared~\cite{cassidy1998}.
Therefore an explicit identification of a vehicle at both measurement 
locations was difficult and made the calculation of the travel-time 
distribution impossible.

\begin{center}
\begin{figure} 
\includegraphics[width=0.9\linewidth]{\DIR/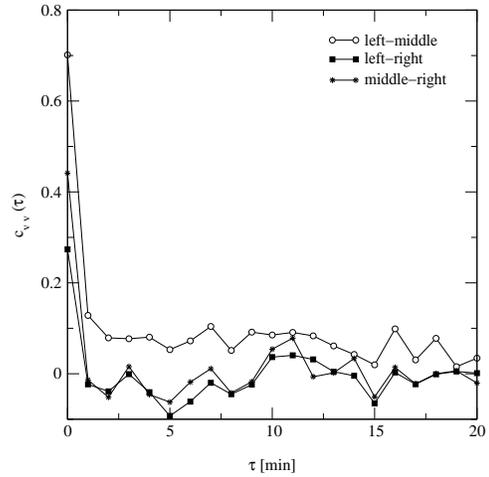}
\caption{Cross-correlation between the minute averaged velocities on the three
lanes on the 04-28-2000 at detector D1 on the A3. Note, that the
velocity time-series was detrended by a linear fit.}
\label{ff_ac3}
\end{figure}
\end{center}

In~\cite{neubert99} it was shown that in free flow traffic the
correlation of the average velocity between different lanes is
negligible.
However, this result is contrasted by our analysis of the velocity
time-series. In Fig.~\ref{ff_ac3} the 
cross-correlation between the detrended average velocities on lane $i$
and lane $j$ is depicted. 
Obviously, there are strong correlations between the lanes, especially
neighboring lanes are more correlated
than the left with the right lane. 
Lane changes lead to a strong coupling between the lanes while the
right lane is somehow decoupled from the other two due to the
large amount of trucks.
These lane changes are forced by the absence of a speed limit that
may explain the differences to the results of~\cite{neubert99} where a
speed limit was in effect.
As a result, daily fluctuations of the mean
velocity and therefore of the flow are served first by the left and
the middle lane while the flow of the right lane does not change
significantly in the free flow regime.



\section{Synchronized flow}

\begin{center}
\begin{figure}[hbt]
\includegraphics[width=0.9\linewidth]{\DIR/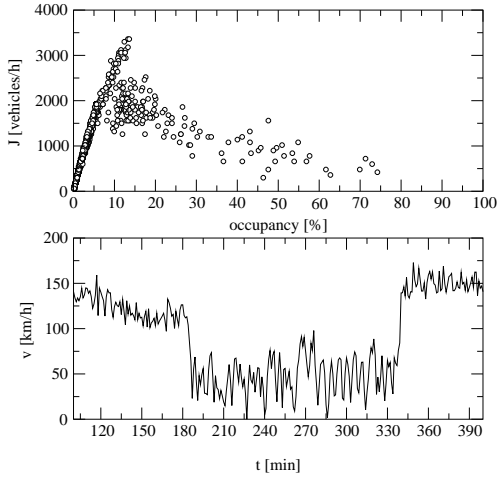}
\caption{Fundamental diagram and time-series of the velocity on
the 05-11-2000 of the left lane at detector D1 on the A3.}
\label{s_fundi}
\end{figure}
\end{center}

Synchronized traffic can be characterized by a large variance in flow
and density measurements and a velocity that is considerably lower
than in free flow traffic while the flow can show large
values (Fig.~\ref{s_fundi}). 
Although the minute averaged velocity is larger compared
to a wide jam, the analysis of the single-vehicle data reveals the
existence of vehicles with velocities smaller than $10$ km/h, that is
vehicles at rest were measured~\cite{Kerner99}. 

\begin{center}
\begin{figure}[hbt]
\includegraphics[width=0.9\linewidth]{\DIR/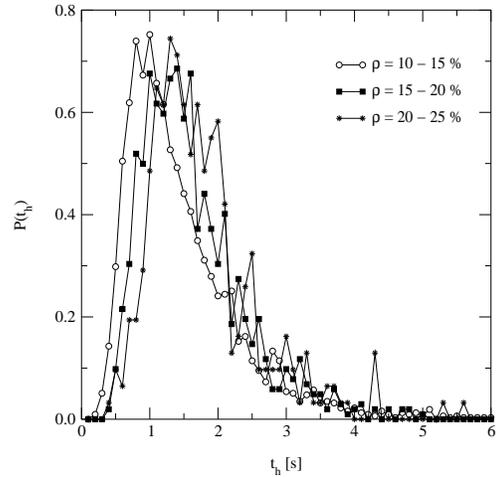}
\caption{Time-headway distribution in the synchronized regime for all days
of the left lane at detector D1 on the A3.}
\label{s_th}
\end{figure}
\end{center}

In contrast to free flow, the time-headway distribution shows a non-trivial
dependence on the density (Fig.~\ref{s_th}). While the maximum of the
distribution is 
shifted to larger values ($\approx 1.3$ s) and the variance is
reduced, small time-headways do still exist. The probability of these
small time-headways decreases with increasing density whereas the
distribution in free flow can simply be rescaled by the density for
small times. Note, that the distribution
was calculated by the consideration of all days of the data set with
synchronized traffic.

\begin{center}
\begin{figure} 
\includegraphics[width=0.9\linewidth]{\DIR/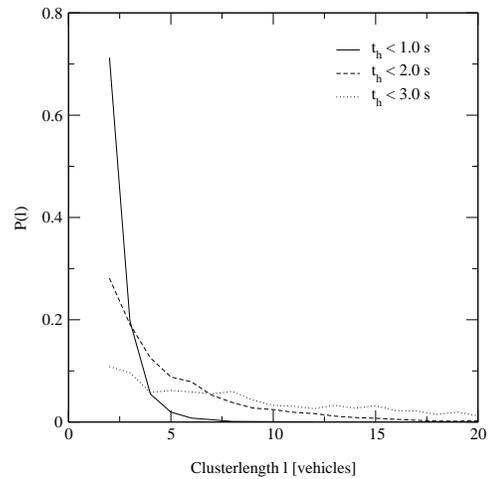}
\caption{Distribution of the cluster length on the left lane for different
time-headways in the synchronized regime.}
\label{s_cluster1}
\end{figure}
\end{center}

Compared to free flow traffic the length of clusters with small
time-headways is decreased significantly (Fig.~\ref{s_cluster1}). 
The increased density in
the synchronized regime is therefore responsible for the dissolution
of the large platoons of the free flow regime. Moreover, because the
velocity of the cars is now determined by the available space rather than
by the desired velocity, the probability of cars piling up behind a
slow car is drastically reduced. 
Now, safety reasons require 
larger distances to the preceding vehicle in order to prevent
braking maneuvers. 
Note, that large clusters with time-headways larger than $2$ s between
the vehicles are
obtained since the mean time-headway has the same order of magnitude.

Nevertheless, strong correlations of the velocity of the vehicles can
be measured 
while the time-headway and the distance-headway
are only weakly correlated~\cite{neubert99}.
Thus, although the velocity of succeeding vehicles is strongly
synchronized their gaps can vary considerably.
As a result, platoons cannot be recognized.

\begin{center}
\begin{figure}[hbt]
\includegraphics[width=0.9\linewidth]{\DIR/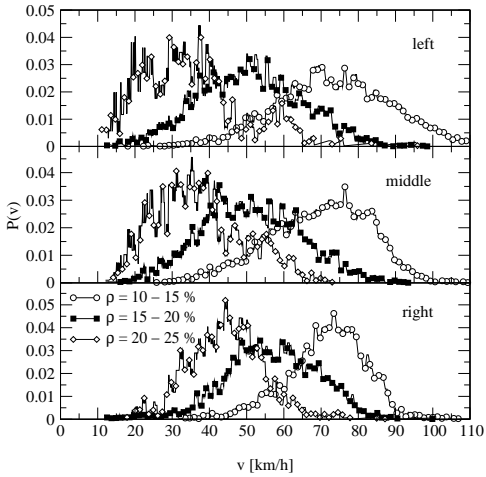}
\caption{Corrected distribution of the velocity in the synchronized regime.}
\label{vdis_synchro}
\end{figure}
\end{center}

Also for the velocity distributions we found significant differences 
to free flow traffic. In synchronized traffic small velocities have,
compared to a Gaussian distribution, a much higher weight
(Fig.~\ref{vdis_synchro}). The 
form of the velocity-distribution is probably a consequence of the 
accelerating and braking vehicles. Our result is of special importance 
for some macroscopic models of traffic flow \cite{Helbing2000},
 that assume Gaussian distributions 
of speeds,  clearly incompatible with our empirical findings.

As shown in~\cite{neubert99} the velocity-headway curve gives
information about the average velocity of the vehicles at a given
headway.

In the free flow regime large velocities are reached even for
very small distances but in synchronized traffic the asymptotic
velocity is reduced considerably (Fig.~\ref{s_ov}).
Again, the velocity-headway curve was obtained by averaging over all
days with synchronized traffic.
Since the vehicles drive slower than the gap allows the gap usage is
only suboptimal.
Moreover, the velocity of a vehicle at a given distance depends 
strongly on the traffic density. The larger the density the lower the 
velocity which allows the driver to anticipate breakdowns.

The transition from free flow to synchronized traffic is characterized
by the synchronization of the speed on different lanes and a sharp
drop of the velocity. 
This synchronization is a result of lane changes in order to
align the density of the lanes~\cite{kerner2001,zweispur} rather than a
consequence of the difference between the maximum velocities of the
vehicles.

\begin{center}
\begin{figure}[hbt]
\includegraphics[width=0.9\linewidth]{\DIR/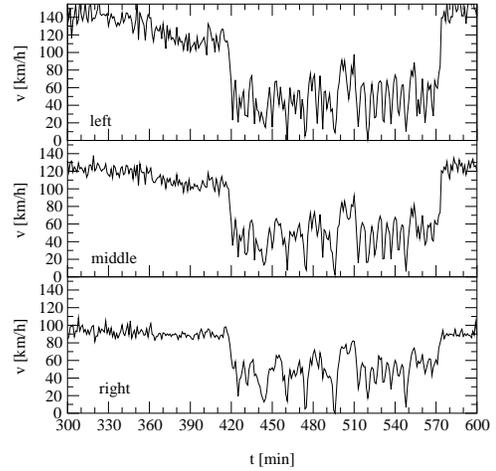}
\caption{Velocity time-series at detector D1 on the A3 on the
05-11-2000 for all three lanes.} 
\label{s_v}
\end{figure}
\end{center}

\begin{center}
\begin{figure} 
\includegraphics[width=0.9\linewidth]{\DIR/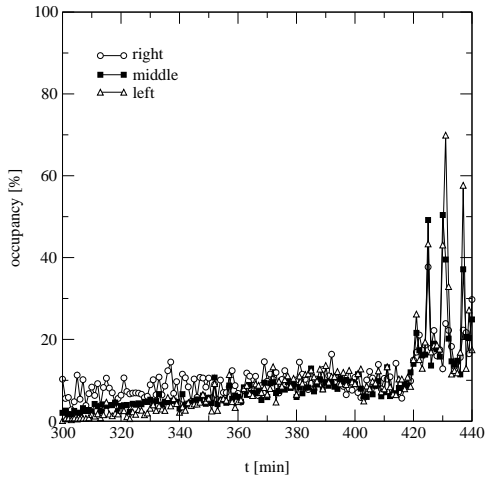}
\caption{Occupancy time-series of the three
lanes at detector D1 on the A3 on the 05-11-2000.}
\label{s_rho}
\end{figure}
\end{center}

In Fig.~\ref{s_v} the time-series of the velocity measurements for the three
lanes is shown. 
A slow decrease of the average velocity on the left and the middle
lane can be seen until a sharp drop of the velocity occurs. 
Obviously, the breakdown happens at the same time for all three lanes.
In contrast to the velocity, the density difference between the lanes
decreases continuously and completely vanishes at the transition
(Fig.~\ref{s_rho}).  
The synchronization of the lanes therefore happens continuously.

The calculation of the cross-correlation between the velocities, the densities
and the flows on the three lanes reveals that the lanes are strongly
correlated, e.g., synchronized. 
As a consequence of this synchronization  
the distribution of the velocity, the distance-headway
and the 
time-headway are nearly the same for each lane.
Thus the basic driving
behavior in synchronized traffic is determined neither by the density nor
by the lane but depends only on the traffic state.

\begin{figure} 
\includegraphics[width=0.9\linewidth]{\DIR/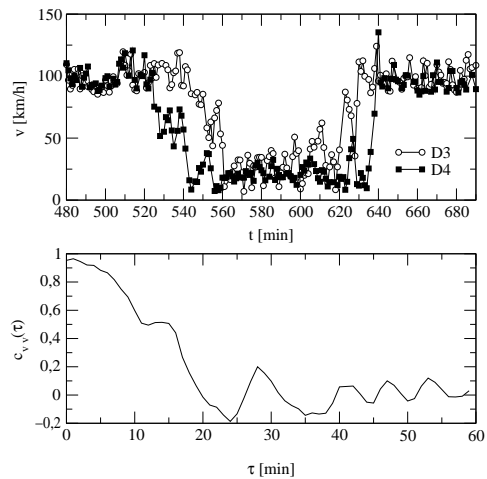}
\caption{Cross-correlation (bottom) between the velocity time-series (top) of
synchronized traffic of the middle lane on the 02-20-2001 at the
detectors D3 and D4 on the A1.}
\label{s_spatialcorr1}
\end{figure}

Again, the cross-correlation analysis of the velocity time-series between
the detectors D3 and D4 on the A1 allows the identification of moving
structures (Fig.~\ref{s_spatialcorr1}).  
The difference in the life-time of the synchronized states at each
location suggests the existence of a bottleneck far upstream of D4
where a transition from free flow to synchronized traffic has 
occurred.

 As an indication for the stochasticity of the jam formation
mechanism 
velocity fluctuations increase with the distance to the 
bottleneck, finally leading to a pinch region with small jams. The
propagation of these small jams leads to a small decay of the
cross-correlations of the order of magnitude of $10$ min which is
about the period of the jam emergence at one measurement location.
However, the upstream velocity of these small jams cannot be
determined due to their continuous excitation and extinction.



\section{Wide moving jams}

\begin{center}
\begin{figure} 
\includegraphics[width=0.9\linewidth]{\DIR/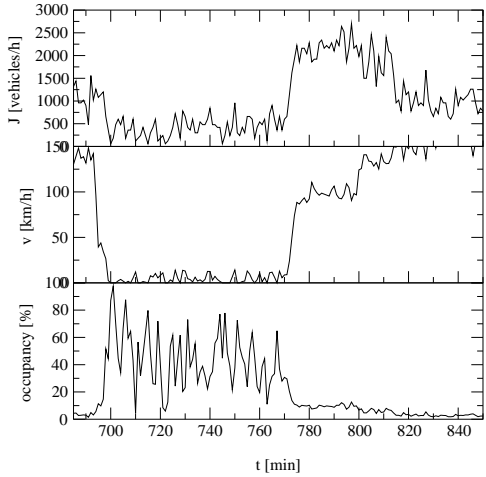}
\caption{Time-series of the flow, the velocity and the occupancy of the
left lane on the A3 at detector D2 on the 04-15-2000.}
\label{wj_fundi2}
\end{figure}
\end{center}

In contrast to free flow and synchronized traffic wide moving jams are
moving structures that are bounded by two fronts where the
vehicle speed changes considerably. The flow and the velocity inside
the wide jam are negligible whereas the density has large values
(Fig.~\ref{wj_fundi2}). 
These large values of the density can be traced back to a large number of
vehicles that passed the detector rather than to a single car at
rest which occupies the detector for a long time.
Shifting the time for $60$ s the minute averaging starts and averaging
over these values in order to obtain a mean density shows no
decrement of the density.

\begin{center}
\begin{figure} 
\includegraphics[width=0.9\linewidth]{\DIR/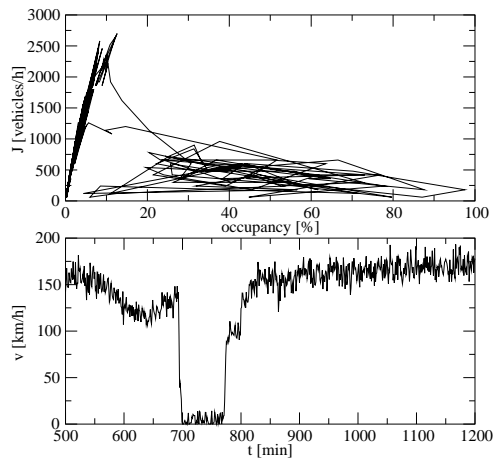}
\caption{Fundamental diagram and velocity time-series of the
left lane on the A3 at detector D2 on the 04-15-2000.}
\label{wj_fundi1}
\end{figure}
\end{center}

In the fundamental diagram the wide jam can be identified by a
triangle structure at large densities and small flows
(Fig.~\ref{wj_fundi1}). 
Two effects have to be taken into account in the interpretation of
the data for wide jams: 1) Wide jams are not compact in the sense that 
cars are piled up bumper-to-bumper, but there can still exist 
relatively large gaps inside. These gaps lead to small values of the 
density and the flow.
2) Standing vehicles can block the detector for a certain time.
This results in large densities but small flows.

The characteristics of a wide jam are the upstream moving velocity and
the outflow from the jam if free flow is formed in the outflow.
As one can see in Fig.~\ref{wj_fundi1} the transition from a wide jam
to free flow is
accompanied by a jump of the velocity to values lower than in
free flow traffic while the density and the flow are considerably
larger (Fig.~\ref{wj_fundi2}) determining $J_{out}$ and
$\rho_{out}$~\cite{kerner2001} to be $2200$ vehicles/h and $10\%$.
In addition, the difference of the outflow from a jam to free flow
traffic decreases from the left to the right lane due to the
decreasing average velocity.

\begin{center}
\begin{figure} 
\includegraphics[width=0.9\linewidth]{\DIR/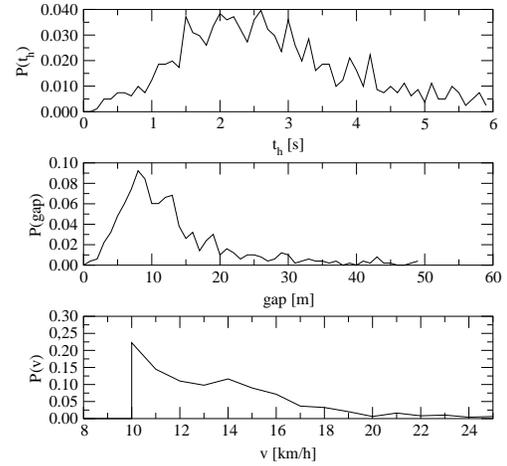}
\caption{Distribution of the time and the distance headway and
the velocity in a wide jam for the left lane on the A3 at detector
D2 on the 04-15-2000 and  at detector D1 on the 05-02-2000.}
\label{wj_dis1}
\end{figure}
\end{center}

Inside the jam the velocity and the flow are negligible.
Since vehicles can only move due to the propagation of holes, the 
velocity inside a wide jam is distributed only in a very
small range (Fig.~\ref{wj_dis1}).

Note, that the cut-off is due to
measurement restrictions, e.g., velocities smaller than $10$ km/h 
are not detected.
As mentioned before, the calculation of the
headways is only correct under the assumption that the velocities of
two consecutive vehicles do not change significantly.
However, relatively large values of the time and the distance headway 
become visible (Fig.~\ref{wj_dis1}) which can be explained by gaps
within the jam. Hence, the wide jam is not compact but 
gaps allow the movement of the vehicles.

\begin{center}
\begin{figure} 
\includegraphics[width=0.9\linewidth]{\DIR/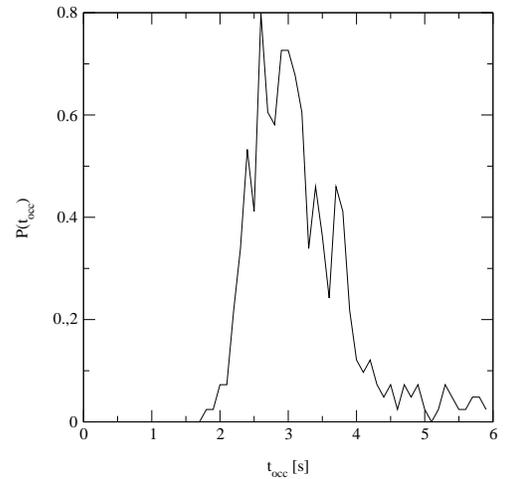}
\caption{Distribution of the time the detector is
occupied by a car with a velocity smaller than $10$ km/h in a wide jam.} 
\label{wj_occdis}
\end{figure}
\end{center}

\begin{center}
\begin{figure} 
\includegraphics[width=0.9\linewidth]{\DIR/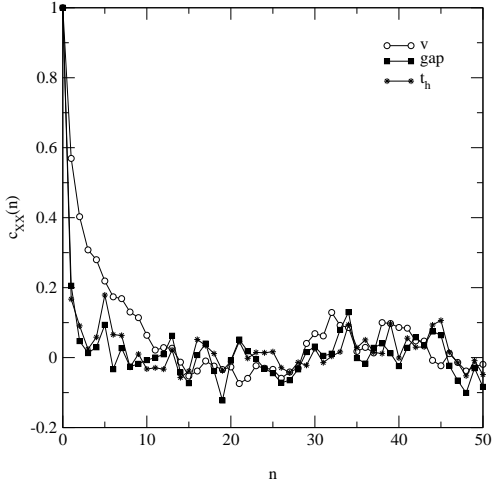}
\caption{Autocorrelation of the single-vehicle data of the
velocity and the time and the distance-headway in a wide jam on the
left lane on the 04-15-2000 at the detector D2 on the A3.}
\label{wj_ac1}
\end{figure}
\end{center}

If vehicles with a velocity smaller than $10$ km/h pass the detector
the velocity is not detected but instead of the length of the vehicle
its time blocking the detector is given. This time, the occupancy
time $\Delta t$, allows to estimate the time, a car at rest needs to
accelerate~\footnote{Since it is possible that the vehicle
has to stop at the detector the assumption of a constant velocity is
no longer valid. Thus, the calculation of the time-headway is not
possible.}. In Fig.~\ref{wj_occdis} the distribution of the occupancy 
time is shown. Obviously, a vehicle needs a minimal time of about
$2$ s to accelerate. 
As a characteristic property of a wide jam~\cite{kernerprl1998} this
time determines (including the density inside the jam) the escape rate
from a jam and thus its outflow.
As a consequence of the gaps inside the jam a stop-and-go pattern
forms that leads to an uncorrelated movement of the
vehicles. However, in contrast to the autocorrelation function of the 
time-headway and the distance-headway the correlation of the
single-vehicle velocity decays slower  (Fig.~\ref{wj_ac1}).

 Thus, weak
correlations 
between the vehicles on very short length scales exist that can be
traced back to the successive acceleration of stopped vehicles.

In analogy to synchronized traffic the 
transition from free flow traffic to wide jams is accompanied by an
alignment of the densities on the different lanes.
In contrast to the transition from free flow to synchronized traffic
the velocity difference between the lanes does not decrease indicating
that the breakdown happens due to a large perturbation.
Nevertheless, the breakdown of the velocity
time-series can be observed at the same time on all lanes which is a
consequence of the synchronization of the lanes.

As the next step we analyze the so-called lane-usage inversion.
The lane-usage inversion is a special phenomenon observed on German
highways~\cite{Sparmann78} and concerns the distribution of the
total flow on the lanes. For low densities nearly all the flow can be
measured on the right lane.

However, above a certain density, the inversion point, most of the
flow can be found on the left lane rather than on the right lane. 
This special observation can be explained by the legal restrictions of
highway traffic in Germany. Since the left lane is the designated fast
lane, vehicles must drive on the right lane where overtaking 
is forbidden.  
As a consequence, 
at larger densities cars keep on moving on the left lane
in order to avoid being trapped behind a slow car on the right lane.
Thus, more and more flow can be found on the left rather than the
right lane. This behavior is contrasted by observations on American
freeways~\cite{Hall88,Chang91}.

\begin{center}
\begin{figure} 
\includegraphics[width=0.9\linewidth]{\DIR/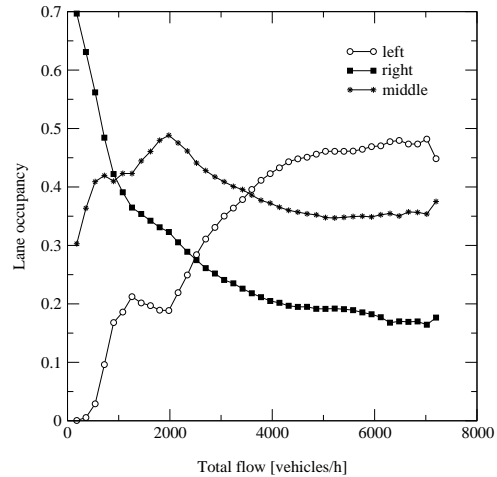}
\caption{Distribution of the total flow on the three lanes.}
\label{wj_flow}
\end{figure}
\end{center}

In Fig.~\ref{wj_flow} the distribution of the flow on the three lanes
averaged for the whole data set of the A3 is depicted. 
In contrast to observations of two-lane traffic, in three-lane traffic
three inversion points are visible. At a flow of about 
$1000$ vehicles/h the weight of the distribution is shifted from the right
to the middle lane. At a total flow of about $2500$ vehicles/h the second
inversion point is reached and more flow can be found on the left than
on the right lane. Finally, at flows larger than $3500$ vehicles/h the
share of the flow is larger on the left than even on the middle lane.

\begin{center}
\begin{figure} 
\includegraphics[width=0.9\linewidth]{\DIR/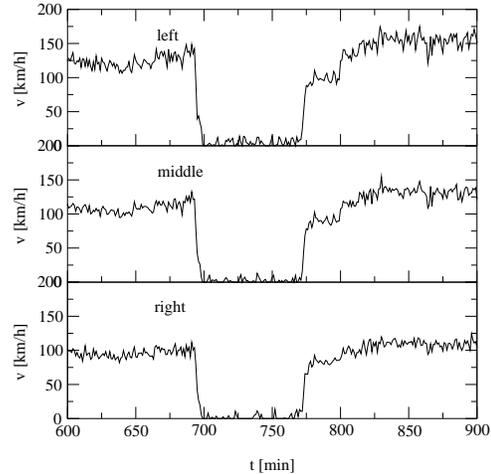}
\caption{Time-series of the velocity for the three lanes on the 04-15-2000 at
detector D2 on the A3.}
\label{wj_v}
\end{figure}
\end{center}

\begin{center}
\begin{figure} 
\includegraphics[width=0.9\linewidth]{\DIR/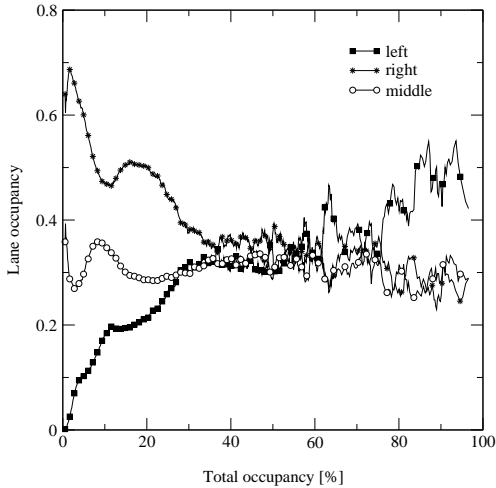}
\caption{Distribution of the total occupancy on the three
lanes. The data were smoothed by a running average.}
\label{wj_occ}
\end{figure}
\end{center}

In contrast to the flow distribution the distribution of the total
occupancy shows no inversion points. As mentioned before, the transitions
from free flow to either synchronized traffic or wide jams is
accompanied by a alignment of the density as shown in Fig.~\ref{wj_occ}.
Right before the
transition the density is evenly distributed on the three lanes. 

For all densities there is a negative density gradient
from the right to the left lane. With increasing density the
difference between the lanes decreases but for no density more
vehicles are on the left than on the right lane.
However, as mentioned in section~\ref{occupancy} occupancy gives the
percentage of the measurement time the detector is covered by
vehicles. Therefore, the slower the vehicles are the larger the
occupancy since slow vehicles occupy the detector a longer time than
fast vehicles.
In order to estimate the spatial density it is necessary to measure
the mean length of the vehicles on the individual lanes. With the
values $L_{\rm{left}} = 4.25$ m, $L_{\rm{middle}} = 4.57$ m and
$L_{\rm{right}} = 8.3$ m the density can be calculated via eqn.(\ref{occden}).

\begin{center}
\begin{figure} 
\includegraphics[width=0.9\linewidth]{\DIR/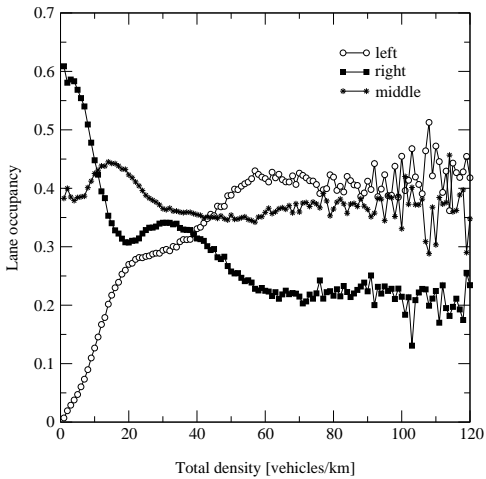}
\caption{Distribution of the total spatial density on the three
lanes.}
\label{laneusage_density}
\end{figure}
\end{center}

As one can see in Fig.~\ref{laneusage_density} there is indeed a
lane-usage inversion. At densities well below a certain point more
vehicles are on the left than on the right and the middle lane, leading
to a negative flow gradient from left to right.
In addition, the cross-over of the density distribution occurs
analogously to the flow distribution at three inversion points. 
As a consequence of the large amount of trucks on the right lane,
the left and the middle lane show nearly the same lane
occupancy which is considerably larger compared to the right
lane. This fact is another indication for the decoupling of the right
from the left and the middle lane.


Thus, the lane usage inversion can be traced back on the one hand to
differences 
of the average velocity between the lanes that lead to larger flows
from right to left and on the other hand to an asymmetric
distribution of the vehicles on the lanes (the flow inversion can also
be observed on highways with speed limit).


\section{Conclusion}

Our analysis of single-vehicle traffic data 
leads to a more 
elaborated microscopic picture of highway traffic.
In the free flow state correlations between the vehicles are
only visible on scales smaller than one minute. The correlations
result from platoons, i.e., clusters of cars driving bumper-to-bumper
with very small time-headways. 
Within these clusters the vehicles are
driving with nearly the same velocity leading to a large 
temporal stability of the platoons. 
With increasing size of the 
platoons we observe a larger distance between the vehicles 
inside the platoon.


In contrast to highways without speed limit,
with speed limit
drivers tend to drive with larger temporal distance, 
i.e., the probability of very small
headways is reduced considerably and 
the maximum of the time-headway distribution 
is shifted towards larger values. This may indicate that speed-limits 
lead to a reduction of accidents. But even more efficient than that 
would be the reduction of the number of drivers driving with 
time-headways comparable to their reaction time, e.g., by 
equipping cars with devices that adjust automatically the distance
to the leading car. 

Another remarkable effect that can already be observed in the free
flow regime is the synchronization of speeds on the 
different lanes, which was predicted 
by different simulation studies~\cite{zweispur}.  

The velocity distributions, that are important as an input
for macroscopic traffic models~\cite{Helbing2000}, have the expected 
Gaussian shape  
for densities of $5-10\%$. Surprisingly this is not true for 
very low densities. There the distribution functions show a 
slight asymmetry, with an increased weight of large speeds.
  
The synchronized traffic state is known to be characterized by 
increased fluctuations, e.g., of the velocity time series, compared 
to free flow traffic. On a microscopic level these fluctuations 
influence the time-headway distributions on short time scales. The 
short headways are systematically suppressed and the maximum of the 
distributions is shifted towards larger headways. This effect may 
be explained by the tendency of drivers to avoid frequent accelerations
and brakings~\cite{knospe2000}. 
Other consequences of the larger velocity fluctuations are the 
more rapid dissolution of the platoons in synchronized traffic
and the much higher weight of small velocities. The low 
speed part of the velocity distribution is probably governed 
by acceleration and braking vehicles.

Wide jams can be identified by a triangular shape in the fundamental
diagram. Since they are not compact the transportation of holes leads
to small flows at small densities.
The analysis of the time a vehicle at rest covers a detector allows
to estimate its acceleration time. This time determines the outflow 
from a jam and is characteristic for wide jams.
Our result of about $2$ s confirms the value found in~\cite{kernerprl1998}. 

The transition from free flow to either synchronized traffic or wide
moving jams is accompanied by a continuous alignment of the density
on the different lanes. The breakdown of the velocity
happens simultaneously which is a consequence of the synchronization
of the vehicle's speed on the lanes.

Empirical studies reveal that on German two-lane highways the flow is
distributed asymmetrically on the lanes.
This lane usage inversion is a result of the
asymmetric distribution of the vehicles on the lanes and can be found
even on three-lane highways. It
is increased by a negative velocity gradient from the left to the
right lane on highways without speed limit.

Our analysis confirms the basic results of
former empirical studies~\cite{neubert99,Tilch99TGF}.
In particular, the characteristics of the time-headway distribution
and the velocity-distance curve have been found in both data sets. 
The new data allowed to clarify an open question about the
occurrence of peaks in the time-headway distribution \cite{neubert99}
that are most likely an artifact of the internal data handling of
the inductive loops.
The present results show that a peak at a headway of $1.8$ s does not exist.

Finally, our results confirm that the basic driving strategies implemented
in a recent cellular automaton traffic model~\cite{knospe2000} match 
quite well the empirical situation. Of special importance are 
anticipation effects, but also the reduced gap acceptance in synchronized 
traffic. We also want to stress the fact that the different 
traffic states, which have been identified by means of a 
spatio-temporal analysis of time-averaged data, differ equally 
well with respect to their microscopic structure. 
>From our point of view, this coherence between microscopic and
macroscopic empirical analyses  
supports the validity of the classification scheme.

\bigskip

{\bf Acknowledgments}: 
The authors are grateful to the Landesbetrieb Stra\ss enbau NRW
(especially Mr.~Thomas) for
the data support and to the 
Ministry of Economics and Midsize Businesses, Technology and Transport
for the financial support.
We thank Mr.~Wefer from BIANDI Informationssysteme for useful
discussions and his technical advice concerning the inductive loops.
L.~S.~acknowledges support from the Deutsche Forschungsgemeinschaft
under Grant No. SA864/1-2.  


\begin{appendix}
\section{}
\label{fourier}

The main problem in the analysis of empirical time-series are irregular
observation times or gaps in the observations, i.e., data which
are no evenly spaced.
In contrast, the classical definition of the correlation function of
two observables $X$ and $Y$ measured at $N$ points
\[ c_{XY}(\tau) = \frac{1}{N} \sum_{n=1}^{N-\tau} X_nY_{n+\tau}\]
requires evenly spaced data.
There are some approaches for filling the gaps in the time-series, but
nevertheless the original time-series is changed by these methods.
In order to circumvent these problems, the correlation is calculated
in the frequency domain and then transformated back to the time domain.
The correlation theorem~\cite{jenkins1968} states, that the cross
spectrum of two 
functions is the Fourier transform $F$ of its cross-correlation function:
\[ c_{XY} = F^{-1}[F_X(\omega)F_Y^{*}(\omega)]. \]
The Fourier transform of unevenly spaced data $\{ X_n=X(T_n),
n=1,2,\ldots,N\}$ is computed
after~\cite{scargle} and is defined to be:
\[
F_X(\omega) = F_0 \sum \left( A X_n \cos(\omega t_n^{'} )+ i B X_n
\sin(\omega t_n^{'} ) \right) 
\]
where
\[ 
F_0(\omega) = \sqrt{\frac{N}{2}} e^{-i\omega t_1},\qquad
t_n^{'} = t_n-\tau(\omega)
\]
and
\[ A(\omega) = \frac{1}{\sqrt{\sum_n \cos^2(\omega t_n^{'})}}, \quad \quad 
B(\omega) = \frac{1}{\sqrt{\sum_n \sin^2(\omega t_n^{'})}} \]
and
\[ \tau(\omega) = \frac{1}{2 \omega} \tan^{-1} \left ( \frac{\sum_n
\sin(2\omega t_n)}{\sum_n \cos(2\omega t_n)} \right ). \]

$t_1$ is determined by the origin of time.
See~\cite{scargle} for details of the computing the Fourier transform
and for a Fortran code. 
Inversion back to the time domain can be accomplished using the
Fourier transform for evenly spaced frequencies.

\end{appendix}

\bibliographystyle{unsrt}

\end{document}